\documentclass[times,11pt]{article}
\newcommand{\etal}{{\it et al.\ }}
\newcommand{\eg}{{\it e.g.},\,}

\newcommand{\vs}{{\it vs.\ }}

\newcommand{\beq}{\begin{equation}
  \renewcommand{\int}{\intop\limits}
  \renewcommand{\oint}{\ointop\limits}}
\newcommand{\eeq}{\end{equation}}
\newcommand{\beqarr}{\par\begin{minipage}{11cm} \begin{eqnarray*}}
\newcommand{\eeqarr}{\end{eqnarray*} \end{minipage} \hfill 
   \stepcounter{equation}{\rm (\theequation)}\vspace{3mm}\linebreak}
\newcommand{\bdm}{\begin{displaymath}
  \renewcommand{\int}{\intop\limits}
  \renewcommand{\oint}{\ointop\limits}}
\newcommand{\edm}{\end{displaymath}}

\newcommand{\up}[1]{\ifmmode^{\rm #1}\else$^{\rm #1}$\fi}

\newcommand{\uph}{\up{h}}

\newcommand{\arcd}{\ifmmode^{\circ}\else$^{\circ}$\fi}
\newcommand{\arcm}{\ifmmode{'}\else$'$\fi}
\newcommand{\arcs}{\ifmmode{''}\else$''$\fi}

\newcounter{pagefrom}
\newcounter{pageto}
\newcounter{volume}
\newcounter{year}

\newenvironment{Titlepage}{
\vspace*{2cm}
  \begin{center}
}{
  \end{center}\par\vspace{3mm}
}

\newcommand{\Title}[1]{{\large\bf\boldmath #1 \\[3mm] {\footnotesize by} 
\\[3mm]}}

\newcommand{\Author}[2]{{\large\spaceskip 2pt plus 1pt minus 1pt #1}\\[3mm]
   {\small #2}\\[6mm]}

\newcommand{\Received}[1]{}

\newcommand{\Abstract}[2]{{\footnotesize\begin{center}ABSTRACT\end{center}
\vspace{1mm}\par#1\par
\noindent
{\bf Key words:~~}{\it #2}}}

\newcommand{\FigCap}[1]{\footnotesize\par\noindent Fig.\  % 
  \refstepcounter{figure}\thefigure. #1\par}

\newcommand{\TabCap}[2]{\begin{center}\parbox[t]{#1}{\begin{center}
  \small {\spaceskip 2pt plus 1pt minus 1pt T a b l e}
  \refstepcounter{table}\thetable \\[2mm]
  \footnotesize #2 \end{center}}\end{center}}

\newcommand{\TableFont}{\footnotesize}

\newcommand{\MakeTable}[4]{\begin{table}[htb]\TabCap{#2}{#3}
  \begin{center} \TableFont \begin{tabular}{#1} #4 
  \end{tabular}\end{center}\end{table}}

\newcommand{\MakeTableSep}[4]{\begin{table}[p]\TabCap{#2}{#3}
  \begin{center} \TableFont \begin{tabular}{#1} #4 
  \end{tabular}\end{center}\end{table}}

\newenvironment{references}%
{
\footnotesize \frenchspacing

\renewcommand{\AA}{Astron.\ Astrophys.}

\newcommand{\Acta}{Acta Astron.}
\newcommand{\MNRAS}{MNRAS}
\renewcommand{\and}{{\rm and }}
\section{{\rm REFERENCES}}
\sloppy \hyphenpenalty10000
\begin{list}{}{\leftmargin1cm\listparindent-1cm
\itemindent\listparindent\parsep0pt\itemsep0pt}}%
{\end{list}\vspace{2mm}}

\def\TYLDA{~}
\newlength{\DW}
\newcommand{\refitem}[5]{\item[]{#1} #2%
\def\REFARG{#3}\ifx\REFARG\TYLDA\else, {\it#3}\fi
\def\REFARG{#4}\ifx\REFARG\TYLDA\else, {\bf#4}\fi
\def\REFARG{#5}\ifx\REFARG\TYLDA\else, {#5}\fi.}

\newcommand{\Section}[1]{\section{\normalsize\bf#1}}

\newcommand{\Acknow}[1]{\par\vspace{5mm}{\bf Acknowledgements.} #1}
\usepackage{supertabular,lscape,epsfig}
\usepackage{amssymb}
\usepackage{amsmath}
\DeclareSymbolFont{ppa}{OT1}{ppl}{m}{it}
\DeclareMathSymbol{\vv}{\mathalpha}{ppa}{'166}
\begin{document}

%Zwarte naglowki, jeden wiersz
\newcommand{\TabCapp}[2]{\begin{center}\parbox[t]{#1}{\centerline{
  \small {\spaceskip 2pt plus 1pt minus 1pt T a b l e}
  \refstepcounter{table}\thetable}
  \vskip2mm
  \centerline{\footnotesize #2}}
  \vskip3mm
\end{center}}

%Zwarte naglowki, dwa wiersze
\newcommand{\TTabCap}[3]{\begin{center}\parbox[t]{#1}{\centerline{
  \small {\spaceskip 2pt plus 1pt minus 1pt T a b l e}
  \refstepcounter{table}\thetable}
  \vskip2mm
  \centerline{\footnotesize #2}
  \centerline{\footnotesize #3}}
  \vskip1mm
\end{center}}

%Zwarte naglowki, jeden wiersz
\newcommand{\MakeTableSepp}[4]{\begin{table}[p]\TabCapp{#2}{#3}
  \begin{center} \TableFont \begin{tabular}{#1} #4 
  \end{tabular}\end{center}\end{table}}

%Zwarte naglowki, jeden wiersz
\newcommand{\MakeTableee}[4]{\begin{table}[htb]\TabCapp{#2}{#3}
  \begin{center} \TableFont \begin{tabular}{#1} #4
  \end{tabular}\vspace*{-7mm}\end{center}\end{table}}

%Zwarte naglowki, dwa wiersze
\newcommand{\MakeTablee}[5]{\begin{table}[htb]\TTabCap{#2}{#3}{#4}
  \begin{center} \TableFont \begin{tabular}{#1} #5 
  \end{tabular}\end{center}\end{table}}

\newfont{\bb}{ptmbi8t at 12pt}
\newfont{\bbb}{cmbxti10}
\newfont{\bbbb}{cmbxti10 at 9pt}
\newcommand{\uprule}{\rule{0pt}{2.5ex}}
\newcommand{\douprule}{\rule[-2ex]{0pt}{4.5ex}}
\newcommand{\dorule}{\rule[-2ex]{0pt}{2ex}}
\def\thefootnote{\fnsymbol{footnote}}
\begin{Titlepage}
\Title{The All Sky Automated Survey. The Catalog of Variable Stars.
II.~6$^{\rm\bf h}$--12$^{\rm\bf h}$ Quarter of the Southern Hemisphere}
\Author{G.~~P~o~j~m~a~{\'n}~s~k~i}{Warsaw University Observatory,
Al~Ujazdowskie~4, 00-478~Warszawa, Poland\\
e-mail:gp@astrouw.edu.pl}
\end{Titlepage}

\Abstract{This paper describes the second part of the photometric data from
the ${9\arcd\times9\arcd}$ ASAS camera monitoring the whole southern
hemisphere in the {\it V}-band. Preliminary list of variable stars based on
observations obtained since January 2001 is presented. Over 2\,800\,000
stars brighter than ${V=15}$~mag on 18\,000 frames were analyzed and 11357
were found to be variable (2685 eclipsing, 907 regularly pulsating, 521
Mira and 7244 other, mostly SR, IRR and LPV stars). Periodic light curves
have been classified using the automated algorithm, which now takes into
account IRAS infrared fluxes. Basic photometric properties are presented in
the tables and some examples of thumbnail light curves are printed for
reference. All photometric data are available over the INTERNET at {\it
http://www.astrouw.edu.pl/\~{}gp/asas/asas.html} or {\it
http://archive.princeton.edu/\~{}asas.}}{Catalogs
-- Stars: variables: general -- Surveys}

\Section{Introduction}
Following ideas of Paczy{\'n}ski (1997) the All Sky Automated Survey (ASAS,
Pojma{\'n}ski 2002) has been monitoring entire southern sky ($\delta<25\arcd$)
since October 2000. All stars, asteroids and comets brighter than limiting
magnitude ${V\approx14}$~mag (${I\approx13}$~mag) are measured once per
1--3 nights using wide-field (${9\arcd\times9\arcd}$) cameras equipped with
${2048\times2048}$~pixel CCDs.

The prototype ASAS system using small commercial CCD camera
(${768\times512}$), 135~mm f/1.8 telephoto lens and {\it I}-band (Schott
RG-9, 3~mm) filter was used in the years 1997--2000 to monitor 0.7\% of the
sky. It has detected almost 4000 variable stars among 150\,000 observed
objects (Pojma{\'n}ski 2000).

Current ASAS system, located at Las Campanas Observatory (operated by the
Carnegie Institution of Washington), consists of four independent
instruments each equipped with the automated paralactic mount, imaging
optics, standard filter, ${2048\times2048}$ (14~$\mu$m pixels) CCD camera
and a dedicated computer. All four are installed together in a small
automated enclosure.

Two wide-field systems are equipped with the Minolta 200/2.8 APO-G telephoto
lenses and standard {\it V} and {\it I} filters.

Narrow field instrument, ${D=250}$~mm, ${F=750}$~mm, is a Cassegrain
telescope with a three element, Wyne-type field corrector and {\it I}
filter. It gives sharp images (FWHM < 2.2 pixels) in the field of 2\arcd
diameter.

Very-wide-field instrument (${36\arcd\times36\arcd}$) equipped with the
{\it R} filter is intended for investigation of bright, short time-scale
events.

So far only one wide-field instrument ({\it V} filter) achieved the status of
the fully automated system. Its data are acquired, analyzed and immediately
available over the Internet within a few minutes after observation. Data
from other instruments are also processed on-line but not yet placed into
the final catalog.

Analysis of the {\it V}-band data started just after reasonable amount of
data had been collected. As a result the catalog of variable stars in the
0\uph--6\uph quarter of the southern hemisphere was released (Pojma{\'n}ski
2002, hereafter Paper~I). This paper contains the second part of the
analyzed data -- variable stars located in the fields centered between
6\uph and 12\uph of right ascension.

\Section{Observations and Data Reduction}
For the wide-field instruments the whole sky was divided into 709
fields of which 513 (70\%) with $\delta<+28\arcd$ can be accessed from
Las Campanas. Since new enclosure does not set obscuration limits on
declination also stars with $+2\arcd<\delta<+28\arcd$ are now observed.
However, due to the small number of measurements acquired so far for that
strip, these fields are not yet included in the catalog.

Each night several sky-flat and dark exposures are taken after sunset,
followed by over 150 3-minute target images, creating the raw data stream
of 1.5~GB per night from each instrument.

The data reduction pipeline used for ASAS observations was described in
Paper~I. Basically, we are making simultaneous photometry through five
apertures (2 to 6 pixels in diameter). Each aperture data is processed
separately, so one can use data obtained with the smallest one for the
faint objects and with the largest one for the brightest objects in the
catalog. More sophisticated methods like profile fitting and image
subtraction were tested but did not perform well for highly variable and
undersampled ASAS images.

Reducing ASAS data we try to correct photometry for saturation and bleeding
and some overexposed stars are indeed present in the ASAS Catalog. One must
be aware however, that large systematic errors are expected for stars
brighter than ${V\approx7.5}$~mag.

Astrometry is now based on the ACT catalog and achieves positional accuracy
better than 0.2 pixels ($<3\arcs$).

The zero-point offset of photometry is based on the Hipparcos (Perryman
\etal 1997) data. A few hundred Hipparcos stars are usually located in
each ${9\arcd\times9\arcd}$ field and we use them for precise offset
calibration. Due to effects caused by non-perfect flat-fielding, lack of
color terms in transformation and blending of stars, systematic errors
as large as several tens of magnitude could be observed for particular
stars. Differential accuracy is much better, reaching 0.01~mag for bright
stars.
\vspace*{-7pt}
\Section{Variability Search}
\vspace*{-3pt}
Data analyzed in this paper cover almost three years (since 2001) of
observation, but variability detection software was run at the end of 2002,
so many stars were not yet tested due to the insufficient number of
measurements. We plan to repeat variability search on the full data set
after releasing the fourth part of this catalog.
  
Variability analysis was similar to that performed for the ASAS-2 data
(Pojma{\'n}ski 2000).  First, light curves for each star in the field were
extracted and median magnitude and dispersion were calculated (in each
aperture separately).  Stars with high dispersion were selected for further
analysis. This obviously disabled detection of low-amplitude variables, so
Fourier-based detection algorithms on all stars having sufficient number of
good measurements are planned.

For each suspected star AoV period analysis (Schwarzenberg-Czerny 1989) was
performed. Stars with statistics value larger than 10.0 were accepted. The
other stars were subject to long-term variability tests: variance analysis
in variable-length bins and trend analysis (average number of consecutive
observations showing the same direction of brightness change).

Over 40\,000 stars of 2\,800\,000 passed initial selection criteria and had
to be inspected further. About 12\,000 passed visual inspection. In
comparison to the previous work acceptance rate was much higher (30\% \vs
15\%) thanks to longer baseline of observations (more long-term variables
were detected) and more numerous measurements (more periodic variables
confirmed). Fraction of detected variable stars (0.4\%) is still much less
than that obtained by ASAS-2 (2.2\%).

Cross identification of variable stars detected by ASAS-2 in {\it I} filter
and ASAS-3 in {\it V} filter in the overlapping fields revealed that less
than 20\% of the stars have been detected by both instruments. This
indicates that ASAS survey is still far from completeness, even for stars
substantially brighter than its limiting magnitude.
\vspace*{-7pt}
\Section{Variability Classification}
\vspace*{-3pt}
Trying to provide preliminary variability classification we followed the
strategy described in Paper~I.

First, all stars that passed verification process were divided into two
groups: strictly periodic and less regular ones, using simple filter
detecting difference between the actual light curve and averaged (folded)
one.

Automated classification scheme that uses several parameters of the light
curve (period, amplitude, Fourier coefficients) was then applied. Stars
were divided into three eclipsing classes: EC (contact or almost contact
configurations), ED (detached binaries), ESD (remaining semi-detached
systems) and six pulsating: DSCT ($\delta$~Sct), RRAB, RRC (RR~Lyr),
DCEP$_{\rm FU}$, DCEP$_{\rm FO}$ (Cepheids in fundamental and first
overtone mode), M (Mira stars). All other stars (SR, IRR, CV, LBV, Novae,
many multi-periodic, etc.) were classified as MISC.

Classification of variable stars based on their light curve shapes and
period may often lead to errors. For example, many  SR stars tend to fall
into the DCEP class. Without external information (\eg color) it is not
possible to recognize them properly. Infrared data available from IRAS
catalogs (IRAS 1988) can help in such cases.

In Fig.~1 we plotted {\it V} magnitudes \vs ${-2.5\log F_{12}}$ (IRAS
12 micron flux) for DCEP and MIRA stars selected from GCVS (Kholopov \etal
1985). Although the scatter due to the chance epoch of IRAS measurements is
large, distinction between two groups is clear. The same is true for SR
and L variables from the GCVS. With only a few exceptions (blue and yellow
giants and supergiants) all show large IR fluxes and occupy the same region
on the diagram as Mira stars.
\begin{figure}[htb]
\vglue4mm
\hglue-6mm{\includegraphics[width=13.5cm, bb=10 40 730 550]{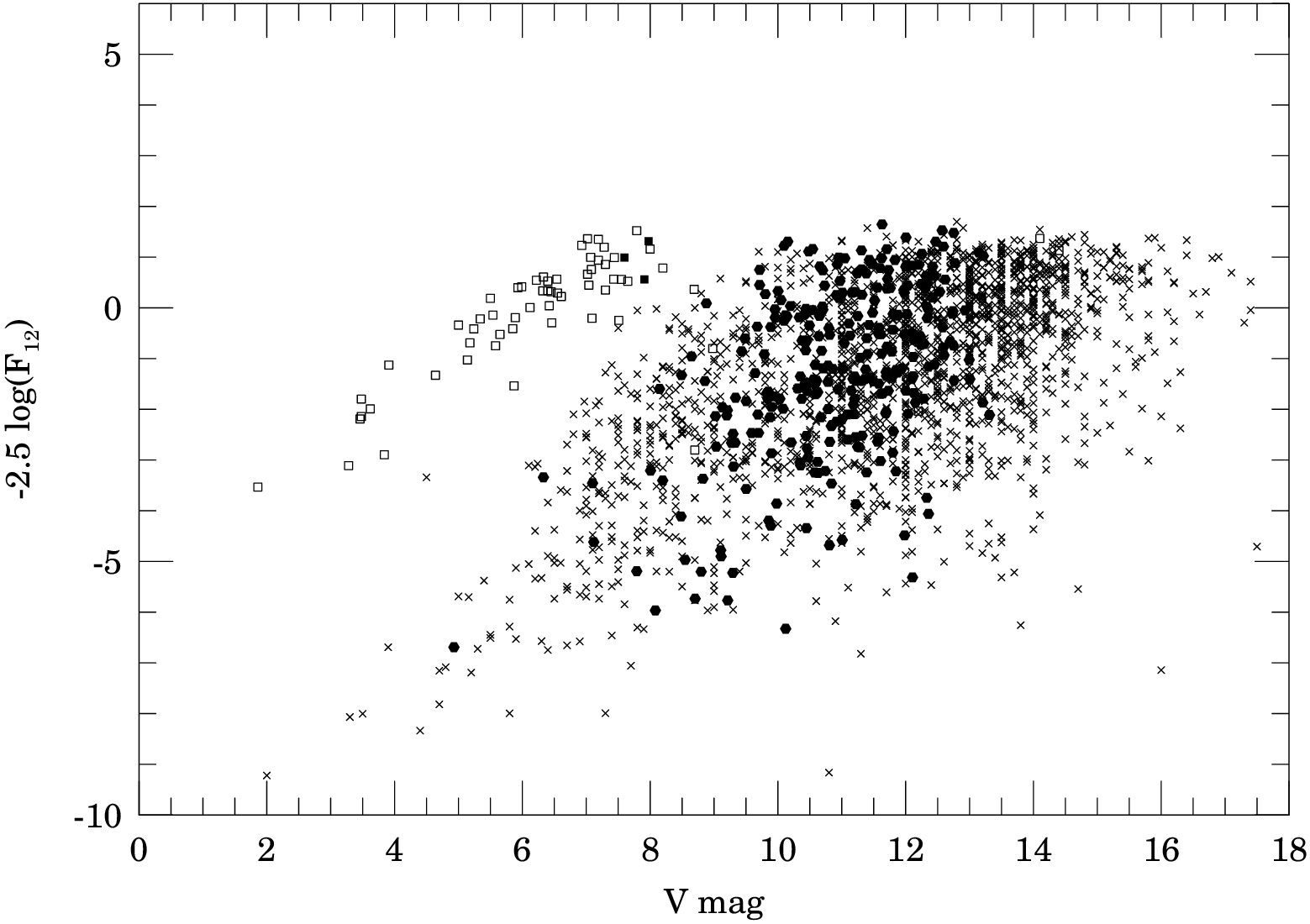}}
\vglue4mm
\FigCap{IRAS 12 $\mu$m fluxes \vs $V_{\rm max}$ magnitudes for Cepheids
and Mira variables from GCVS (open squares and
crosses, respectively) and ASAS (filled
squares and dots, respectively).}
\end{figure}

Plotting GCVS data in Fig.~1 we came across some doubtful cases:
\begin{itemize}
\item{DN Ara (CEP: ${P=82}$~days) is ASAS~172024-6246.0 (SR,
${P=78.5}$ days)}
\item{V1610 Sgr (CEP:) is ASAS~182720-2709.5 (IRR:)}
\item{V686 Cen (CEP, ${P=69.30}$~days) is ASAS~113137-6103.4 (SR,
${P=67.90}$ days)}
\item{DW Mus (CEP: ${P=1.0022}$~days) is ASAS~130750-6853.5 (MIRA,
${P=347}$ days)}
\item{EN TrA (CEP:) is known RVTau star (\eg Van Winckel \etal 1999).}
\end{itemize}
  
In Fig.~1 we also plotted ASAS variables. Only several Cepheids are present
since most of those having IRAS fluxes are brighter than our saturation
limit (${V\approx8}$ mag), and were excluded from the present search. Many
low amplitude SR stars showing large infrared fluxes were initially
classified as DCEP stars. For numerous fainter stars not detected by IRAS
confusion between DCEP and SR is still possible.

\Section{The Catalog}
Current list of candidate variable stars in the second quarter of the
Southern Hemisphere contains 11\,357 stars.  For each star the following data
are provided: ASAS identification ID (coded from the star's
$\alpha_{2000}$ and $\delta_{2000}$ in the form: hhmmss-ddmm.m), period
$P$ in days (or characteristic time scale of variation for irregular
objects), $T_0$ -- epoch of minimum (for eclipsing stars) or maximum (for
pulsating) brightness, $V_{\rm max}$ -- brightness at maximum, $\Delta V$
-- amplitude of variation, Type -- one of the predefined classes: DSCT,
RRC, RRAB, DCEP$_{\rm FU}$, DCEP$_{\rm FO}$, M and MISC.

Stars classified as MISC are mostly semi-regular and irregular
variables detected by our algorithm and objects excluded from
other classes after visual inspection. Original classification was often
appended after MISC keyword. 1866 objects other than MISC have multiple
classification. For 900 cases this is exclusively due to EC/ESD or
ED/ESD confusion, but for 228 other this is a more serious EC/RRC or
EC/DSCT double classification. Quite often visual inspection helped to
remove such degeneracy.

\MakeTable{|l|r|r|r|}{8cm}{Number of various types of detected variable stars}
{\hline
\multicolumn{1}{|c|}{Type} & 
\multicolumn{1}{c|}{$N$} & 
\multicolumn{1}{c|}{Type} & 
\multicolumn{1}{c|}{$N$}\\
\hline
DCEP$_{\rm FU}$	& 239	& EC	& 1317 \\
DCEP$_{\rm FO}$	& 156	& ESD	& 845 \\
DSCT		& 233	& ED	& 495 \\
RRAB		& 189	& M	& 521 \\	
RRC		& 122 	& MISC& 7244 \\
\hline}

Search for detected variable stars in the GCVS catalog revealed about
1050 matches within 3\arcm  radius (60\% of GCVS objects with
$V_{max}<13.5$ in this region). 150 poor matches (distance larger than
30\arcs) were individually inspected.

Table~1 summarizes our classification effort and Table~2 contains a
compact version of the catalog. Only three columns are listed for each
star: identification ID, $V$, and $\Delta V$.  Column ID also contains
some flags -- ":" if classification was uncertain, "?" if multiple classes
were assigned (objects were grouped in the table according to the highest
rank assignment), "v" if SIMBAD lists a star to be variable.

Appendix shows exemplary light curves. Only ID is given for each. For
periodic variables phase in the range ($-0.1$--$2.1$) is plotted along the
$x$-axis, while for Mira's and miscellaneous stars -- HJD in the range
(2451800--2453000). Larger ticks on the $y$-axis always mark 1~mag intervals
and vertical span is never smaller than 1 mag.

The full catalog of variables observed by the ASAS system, containing more
classification details, as well as complete data for the light curves, is
available over the INTERNET:\\ 
\centerline{\it http://www.astrouw.edu.pl/\~{}gp/asas/asas.html} 

or

\centerline{\it http://archive.princeton.edu/\~{}asas}

\Section{Conclusions}
We presented preliminary results of the search for variable stars in
the Southern Hemisphere for right ascension between 6\uph--12\uph. Over
11\,000 variable stars were found among 2\,800\,000 stars brighter than
${V\approx 14}$~mag. Comparison to the ASAS-2 data suggests that our data
are still incomplete and total number of variable stars should increase
substantially when more data are analyzed.

Most of detected variable stars belong to the MISC class, since with
almost two-years long baseline we were able to recognize many semi-regular
and irregular variables.

Our automated classification algorithm, described in Paper~I, was
supplemented with criteria taking into account IRAS fluxes. This helps
discriminating between some SR and DCEP stars. However more research in
this area is still required.

\renewcommand{\arraystretch}{.97}
\renewcommand{\TableFont}{\tiny}

\Acknow{This project was made possible by a generous gift from Mr.\ William
Golden to Dr.\ Bohdan Paczy{\'n}ski, and funds from Princeton University. It is
a great pleasure to thank Dr.\ B.\ Paczy{\'n}ski for his initiative, interest,
valuable discussions, and funding of this project.  I am indebted to the
OGLE collaboration for the use of facilities of the Warsaw telescope at
LCO, for their permanent support and maintenance of the ASAS
instrumentation, and to the Observatories of the Carnegie Institution of
Washington for providing the excellent site for the observations.  This
research has made use of the SIMBAD database, operated at CDS, Strasbourg,
France. This work was supported by the Polish KBN 2P03D02024 grant.}
\input{vartab.tex}

\begin{figure}[p]
\vglue-3mm
\centerline{\bf Appendix}
\vskip1mm
\centerline{\bf ASAS Atlas of Variable Stars. ${\bf 6^{\bf h}{-}12^{\bf h}}$ 
Quarter of the Southern Hemisphere} 

\centerline{\small Only several light curves of each type are printed. 
Full Atlas is available over the {\sc Internet}:}
\centerline{\it http://www.astrouw.edu.pl/\~{}gp/asas/appendix6.ps.gz}
\vskip20mm
\centerline{Stars classified as EC}
\vskip4.5cm
\centerline{Stars classified as ESD}
\vskip4.5cm
\centerline{Stars classified as ED}
\vskip-10.7cm
\centerline{\includegraphics[bb=30 70 400 495, width=13cm]{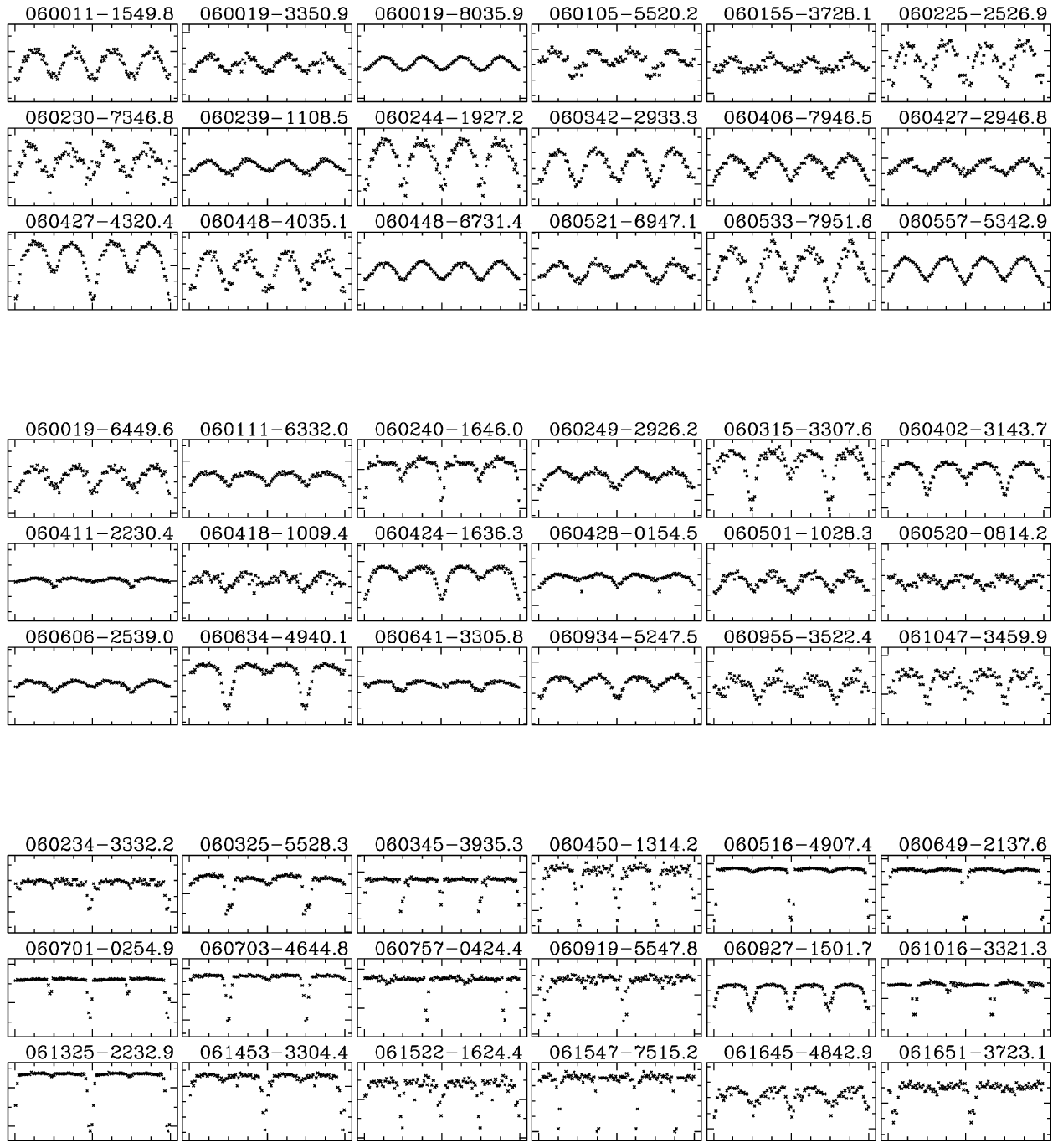}}
\end{figure}
\begin{figure}[p]
\vskip-2mm
\centerline{Stars classified as DSCT}
\vskip4.5cm
\centerline{Stars classified as RRC}
\vskip4.5cm
\centerline{Stars classified as RRAB}
\vskip4.5cm
\centerline{Stars classified as DCEP-FU}
\vskip-14.8cm
\centerline{\includegraphics[bb=30 30 405 610, width=13cm]{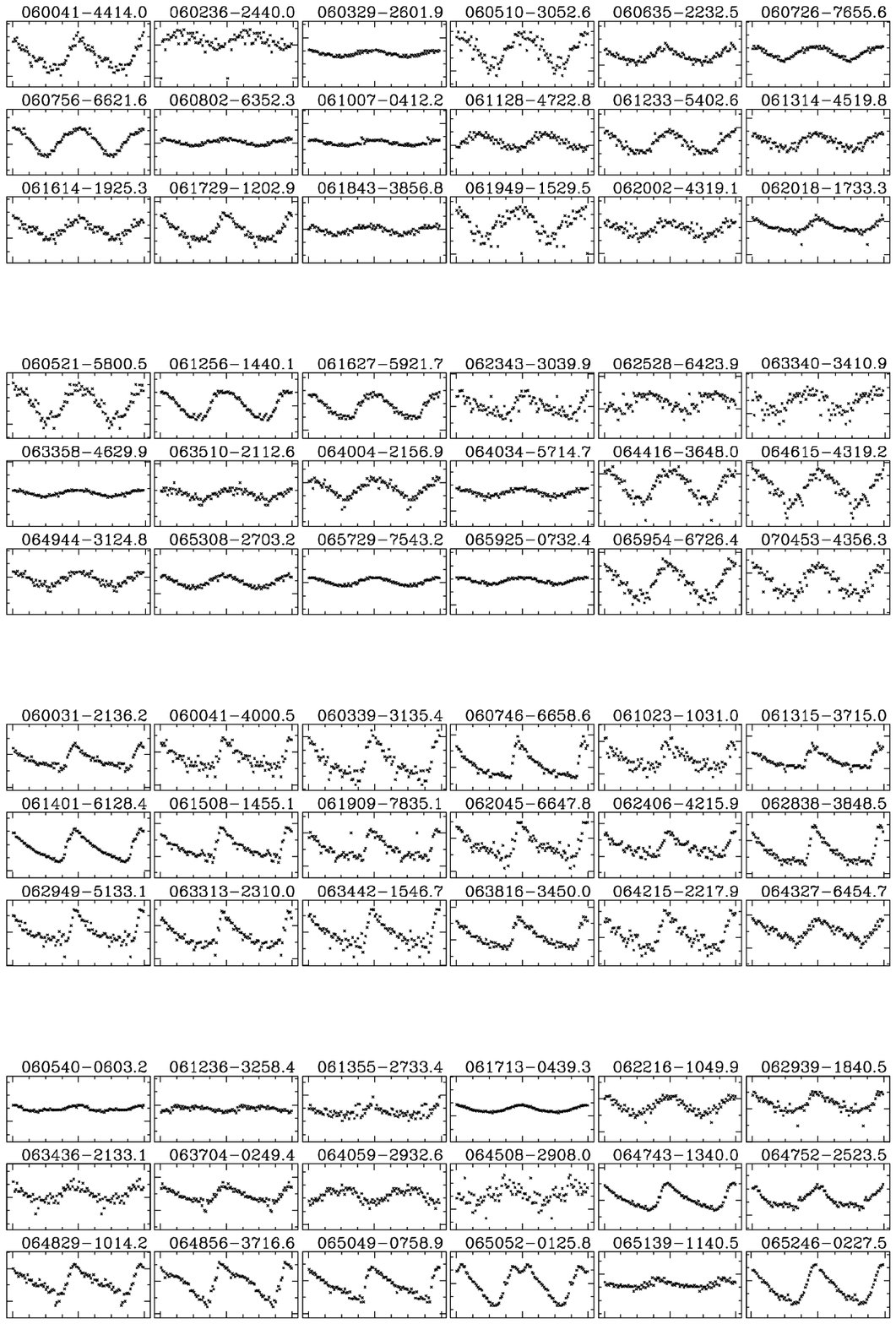}}
\end{figure}

\begin{figure}[p]
\vskip2mm
\centerline{Stars classified as DCEP-FO}
\vskip4.6cm
\centerline{Stars classified as M}
\vskip4.6cm
\centerline{Stars classified as MISC}
\vskip4.6cm
\vskip-14.4cm
\centerline{\includegraphics[bb=30 65 405 610, width=13cm]{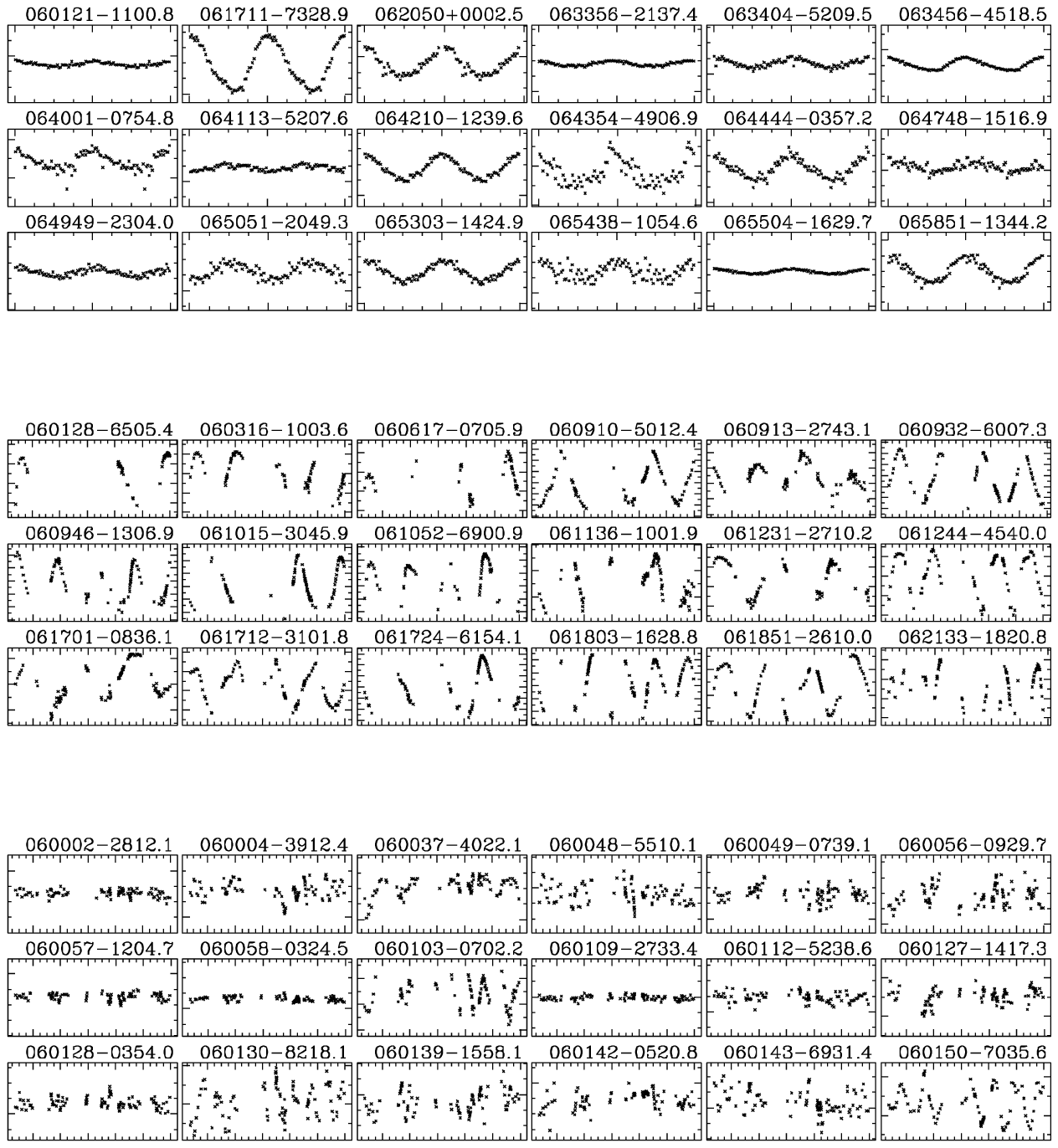}}
\end{figure}
\end{document}